\documentclass{sigchi}

\usepackage{balance}  
\usepackage{graphics} 
\usepackage{times}    
\usepackage{url}      
\usepackage{graphicx,url}
\usepackage{array}
\usepackage{multirow}
\usepackage{bm}
\usepackage{hhline}
\usepackage{tabu}
\usepackage[T1]{fontenc}
\usepackage{times}
\usepackage{helvet}
\usepackage{courier}

\usepackage[show]{chato-notes} 

\usepackage{datenumber}
\usepackage[pdftex]{hyperref}

\makeatletter
\def\url@leostyle{%
  \@ifundefined{selectfont}{\def\UrlFont{\sf}}{\def\UrlFont{\small\bf\ttfamily}}}
\makeatother
\def\pprw{8.5in}
\def\pprh{11in}

\definecolor{linkColor}{RGB}{6,125,233}
\hypersetup{%
  pdftitle={SIGCHI Conference Proceedings Format},
  pdfauthor={LaTeX},
  pdfkeywords={SIGCHI, proceedings, archival format},
  bookmarksnumbered,
  pdfstartview={FitH},
  colorlinks,
  citecolor=black,
  filecolor=black,
  linkcolor=black,
  urlcolor=linkColor,
  breaklinks=true,
}

\newcommand{\mfp}{MFP}
\newcommand{\hashmfp}{{\#}myfitnesspal}

\begin{document}

\title{Persistent Sharing of Fitness App Status on Twitter
\thanks{The majority of the work was done while the first author was at Qatar Computing Research Institute}
}
\numberofauthors{2}
\author{
\alignauthor Kunwoo Park\\
	\affaddr{KAIST}\\
	\affaddr{South Korea}\\
	\email{kw.park@kaist.ac.kr}\\
\alignauthor Ingmar Weber\\
	\affaddr{QCRI}\\
	\affaddr{Qatar}\\
	\email{iweber@qf.org.qa}\\
\alignauthor Meeyoung Cha\\
	\affaddr{KAIST}\\
	\affaddr{South Korea}\\
	\email{meeyoungcha@kaist.ac.kr}\\
\alignauthor Chul Lee\\
    \affaddr{MyFitnessPal}\\
    \affaddr{United States}\\
    \email{clee@myfitnesspal.com}\\
}

\CopyrightYear{2016}

\maketitle

\begin{abstract}

As the world becomes more digitized and interconnected, information that was once considered to be private such as one's health status is now being shared publicly. To understand this new phenomenon better, it is crucial to study what types of health information are being shared on social media and why, as well as by whom. In this paper, we study the traits of users who share their personal health and fitness related information on social media by analyzing fitness status updates that MyFitnessPal users have shared via Twitter. We investigate how certain features like user profile, fitness activity, and fitness network in social media can potentially impact the long-term engagement of fitness app users. We also discuss implications of our findings to achieve a better retention of these users and to promote more sharing of their status updates. 
\end{abstract}

\section{Introduction}

Nowadays, one's health status can be easily collected and shared on social media with the prevalence of mobile and wearable devices along with the improvement of their tracking technology~\cite{jmir08}. Popular health and fitness apps such as Endomondo\footnote{http://www.endomondo.com} and Fitbit\footnote{http://www.fitbit.com} can now help people keep tracking of what they eat, when they exercise, and how well they sleep. These new technologies have facilitated trends like the \emph{Quantified Self} movement, which refers to a group of people who diligently record and monitor different aspects of their daily life~\cite{ChoeLLPK14}. 

Fitness apps provide a wide range of different functionalities such as notification, badging, competition, and community support to achieve a better user retention and engagement of their users~\cite{swan2009emerging}. One such functionality that has been widely adopted is so-called \textit{social sharing}, which lets fitness app users share and broadcast their fitness activity through their social networks. Social sharing is believed to be an effective way of boosting overall user engagement especially when users can get a valuable feedback on their fitness activity by sharing their own personal health and fitness status updates through Social Awareness Streams (SAS) on popular platforms like Twitter or Facebook. 

While previous studies have examined why users would share personal health status updates via Twitter or Facebook and how such sharing may incur both positive and negative effects, not much is known about who \emph{would continue to be engaged with} such social sharing activity for the long-run. We believe that it is important to understand characteristics of such long-run~(or \textit{persistent}) social sharing since it might actually prolong the engagement of fitness app users with physical activity. 
Therefore, the main question that we intend to ask in this paper is \textbf{\textit{what are characteristics of users who share health and fitness-related activity on social media outlets over an extended period of time?}}  Investigating which features are correlated with a persistent sharing of personal health and fitness status updates on SAS can provide a valuable insight for fitness app designers to come up with some features that can help achieve the better retention of their users.

\begin{figure}[h]
{
	\centering
	\vspace{-2mm}	
	\includegraphics[width=.99\linewidth]{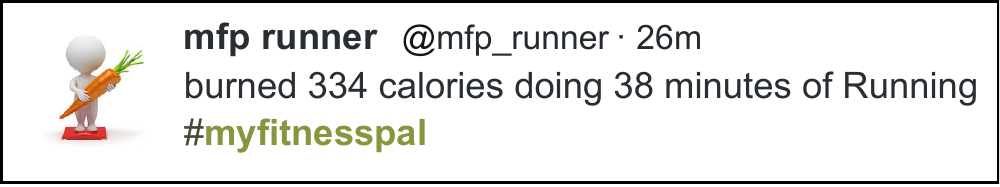}
	\vspace{-1mm}
	\caption{Illustration of Social Sharing on Twitter.}
	\label{fig:mfp_ex}
}
\end{figure}

This paper presents an analysis carried out over an extensive amount of Twitter data generated from one of the most popular fitness apps, MyFitnessPal (MFP for short). {\mfp} users can easily opt-in to automatically share their health and fitness status updates via Twitter (whom we simply call as \textit{social opt-in users}) as illustrated in Figure~\ref{fig:mfp_ex}. These auto-generated stream of posts comprises various types of messages like exercise logs, weight loss updates, and diary updates. We analyze 4.7 million tweets posted by 3,169 {\mfp} users to study which factors are important for the long-term sharing of their fitness status updates.	

Our analysis also provides a valuable opportunity to examine how a wide range of information like Twitter profile information (e.g., followers, following, bio, etc), fitness activity information (e.g., exercise or weight loss records), or fitness network information (e.g., favorites on fitness related tweets, friends network based on fitness interests, etc) can actually affect the persistent engagement of fitness app users. This research further investigates whether certain features can actually predict their long-term engagement.

Our findings can be summarized as follows:

\begin{enumerate}

\vspace*{-1mm}
\item Persistent users tend to use Twitter in a more health-oriented way.

\vspace*{-1mm}
\item Persistent users and short-lived users differ in the types of messages that they share via Twitter. More precisely, sharing of physical activity via Twitter is positively correlated with their long-term engagement.

\vspace*{-1mm}
\item Persistent users and short-lived users differ in the types of networks they form in social media. More precisely, owning a network of friends who are interested in fitness related topics is positively correlated with their long-term engagement.

\vspace*{-1mm}
\item Popularity on Twitter (or frequently receiving favorites on one's tweets) is negatively correlated with the long-term engagement of fitness app users.

\end{enumerate}

We believe that our findings are important to better understand the interplay between fitness apps and social media. We discuss possible implications of these findings as well as their limitations for the rest of the paper.

\section{Related Work}\label{sec:relatedwork}

We review previous research related to health and fitness with a special emphasis on the sharing of health status updates in Social Awareness Streams (SAS).

\subsection{Technological Advances in Health and Fitness}

As information and communication technologies evolve, online tools are increasingly employed to track diet and exercise records that are found to be in such forms like blog postings, forum postings, and weight management applications. 
Several studies have been conducted to efficiently design support tools for weight loss and diet planning. One study aimed to predict the overall success of weight loss by using content from weight loss blogs and found that certain linguistic features had correlation with weight loss~\cite{chung2008predicting}. Other study used publicly available {\mfp} food diary data to study which types of foods are most often associated with weight loss or weight gain~\cite{weber2016dieting}. Researchers have also examined the effectiveness of using support tools during a diet; the role of forums data during the course of a diet was studied by investigating web search logs of nearly 2,000 users 
~\cite{schraefelWAT09}. Dieters progressively narrow their searches to support the progress of their diet on online forums. This study illustrates the overall importance of dieting forums for the maintenance of dieting actions. Other set of researchers found that community interactions are key drivers to make sure that somebody continues to be committed to a diet~\cite{prochaska1997transtheoretical}.

Other researchers investigated different types of social interactions that occur through diet forums and concluded that social conversations would help people find similar buddies to address their concerns and get feedback~\cite{gao2011social}. This means that social factors on diet forums are meaningful to better understand the mechanism for diet. Other study found that users do not only consume information via online forums, but also via social support (e.g., encouragement and motivation)~\cite{Hwang:2011ux}. Other study showed that social networks play an important role for user's engagement in health forums. It is found that user-created groups were socially more active than site-defined groups as these tend to foster more persistent engagement of users in health forums~\cite{vydiswaran2014user}.

Based on these findings, the research community has also been contributing to a better design of effective tools to help people stick with their diet plans to eventually achieve weight loss. A study on a diet portal called \textit{Total Wellbeing Diet} investigated which factors were associated with the user retention on its website~\cite{FreyneSBB012}. It was also found that the usage level during the first week of trial and the number of friends that user has were important factors for her continuous commitment to a diet.  

\subsection{Sharing Health Status Updates in SAS}

It is increasingly common for people to share their health and fitness status updates using Social Awareness Stream (SAS) platforms. Researchers have conducted various studies to understand why users share their health status updates via social network and how their sharing behavior affects their future usage. These studies have demonstrated the importance of user's social network during the course of a diet with the ultimate goal of achieving weight loss.

Some researchers investigated the overall ability of existing social media outlets for health interventions by deploying a Facebook app~\cite{munson2010happier}. They discovered that people tend to adhere to their apps, and are aware of others' privacy concerns because they do not want to spam others' newsfeed. Other researchers investigated why and how people share health information online, by interviewing those who engage with both online health communities and Facebook~\cite{newman2011s}. Participants in these interviews shared their personal health status updates in the pursuit of more emotional support, accountability, motivation, advice, etc. Participants also mentioned that they wanted to manage their projected image to their friends as somebody healthy. Interesting enough, some negative effects were also reported. Privacy concerns kept some participants from uploading their health status updates to Facebook, which would prevent them from receiving all necessary social support. This study suggests the importance of having a group of supportive, but insensitive to privacy concerns, friends who will keep sharing their fitness status updates in SAS. 

A study explored how people would utilize Twitter to share information related to health-promoting physical activity~\cite{kendall2011descriptive}. Relying on Twitter's Streaming API and by filtering exercise-related keywords such as ``elliptical'' and ``muscle-strengthening'', researchers analyzed content from certain tweets to show what and how these were generated. Similarly, other researchers conducted n-gram and qualitative discourse analyses by using public data across different social media outlets~\cite{Chou:2014el}. 

\begin{table*}[t]
  \centering
  \begin{tabular}{llr}
    \textsf{\textbf{Type}} & \textsf{\textbf{Tweet Example}} & \textsf{\textbf{Mean Fraction}}   \\
    \hline
    \textsf{Exercise} & Burned 349 calories doing 30 minutes of running \#myfitnesspal & 0.422   \\
    \textsf{Weight loss} & Lost 5 lbs since her last weigh-in! MFP-USER lost 10 lbs so far. \#myfitnesspal & 0.228   \\
    \textsf{Diary} & Completed her food and exercise diary for 5/05/2014 \#myfitnesspal & 0.338 \\
    \textsf{Blog} & Posted a new blog post SHORT-URL \#myfitnesspal & 0.008 \\
    \textsf{Usage} & MFP-USER has logged in for 295 days in a row! \#myfitnesspal & 0.003\\
  \end{tabular}
  \caption{Observed types for auto-generated {\mfp} tweets and mean fraction of each type across user}
  \label{table:mfp_auto}
\end{table*}

Some researchers also looked into personal health and fitness activities on Twitter~\cite{teodoro2013fitter}. Through a qualitative analysis on health-related tweets, these identified different type of content people had shared on Twitter: Plans and Goals, Achieved, and Acts Avoided. They also interviewed 12 participants to understand their true motivation of sharing updates related to personal health-related activity. Participants reported accountability in sharing health progress reports to their audience. Unlike previous research works that have found certain reluctance in sharing health status on Facebook~\cite{newman2011s}, participants in this particular study were naturally more open to sharing such information publicly. Other researchers investigated which features would help to get more responses on personal Twitter messages, and discovered that user-generated content retrieved way more responses (i.e., favorite and retweet) than automatically generated ones~\cite{epstein2015nobody}.

All above studies have found that the effect of social sharing could vary depending on the type of media (e.g., Twitter, Facebook) and social network. These studies, using a qualitative approach, make important findings about why people share their own personal health information, what type of content gets shared, and how these people are positively or negatively affected by their sharing on SAS. These studies did not investigate whether such positive and negative effects were connected to persistent social sharing, which is critical for fitness business in general. We aim to predict persistent users based on user profile, usage and quality of network.

\textit{RQ. What are the main characteristics of users who share fitness app status updates on social media over a long time period compared to those short-lived users?}

\section{Dataset and Methodology}\label{sec:data}

This section describes MFP and its main features, as well as the methodology for gathering and extracting health and fitness status from tweets. 

\subsection{MyFitnessPal}

\begin{figure}[t]
{
\centering
\includegraphics[width=\linewidth]{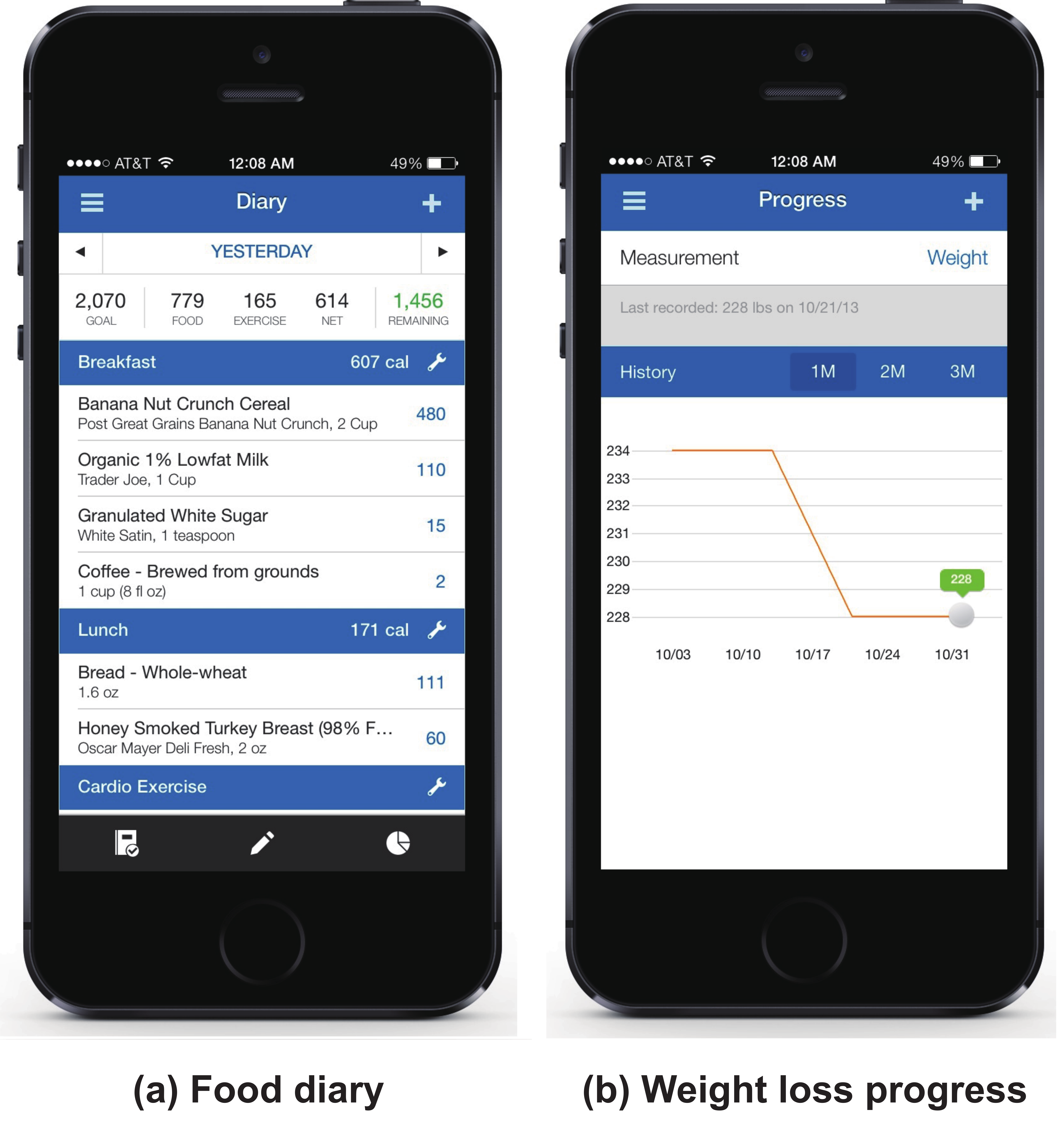}
 \caption{Screenshots of {\mfp} application}
 \label{fig:mfp_description}
}
\end{figure}

{\mfp} is a free health and fitness app that helps people set and achieve personalized health goals through tracking nutrition and physical activity. In 2014, {\mfp} was the most downloaded app in the Google Play Store within the category of health and fitness.\footnote{\url{http://tcrn.ch/1CAph7I}} The {\mfp} app helps users track exercise records, dietary habits, and weight loss to get ``off autopilot'' and get insights that help them make smarter choices and create healthier habits. Upon setting a personal fitness goal, users can visually inspect their fitness and weight loss progress. In Figure~\ref{fig:mfp_description}, we show screenshots of {\mfp} that illustrate how a user of {\mfp} can manage food diaries and track her weight-loss progress using the app. {\mfp} operates its own social network, blogging platform, news feeds, and user forums. {\mfp} also provides an auto-sharing option for certain social media apps like Twitter or Facebook. We refer such auto-sharing option as \textit{social opt-in} in this paper: any user of {\mfp} can opt-in to post status updates, broadcast weight loss, exercise logging or announce the completion of her daily diary through Facebook or to automatically auto-post activity logs and user-generated messages through Twitter using the {\hashmfp} hashtag.

\subsection{Twitter Dataset}

For our analysis, we gathered a large collection of tweets containing the {\hashmfp} hashtag, which we will refer to as ``{\mfp} tweets'' for the rest of the paper. We utilized the Decahose service from GNIP, which provides 10\% of all public tweets, sampled uniformly at random. During a two-month period from April to June in 2014, we retrieved all {\mfp} tweets, corresponding to 6,047 unique users. In addition to the Decahose data, in June 2014, we identified 1,447 additional {\mfp} users who had posted at least one {\mfp} tweet by using the Twitter Streaming API. Six months later, we crawled up to 3,200 public tweets of these users (that include {\hashmfp} tweets as well as general tweets) and all of their following/follower links. Note that we conduct such additional data collection since data collected through Decahose and Streaming API was only limited to sampled tweets. After excluding users who did not post any tweet over the recent three months to ensure users are active on Twitter, the final data set comprises 10,443,988 tweets, 3,188,763 outbound links (following) and 2,766,024 inbound links (followers) of 7,494 users of {\mfp}.  

We limit our observation to a subset of English speaking users (to ensure consistency in the choice of cohort) who own no more than 10,000 followers (to ensure that those micro-celebrities or evangelists who exhibit distinct behaviors are disregarded)~\cite{romero2010directed}. Furthermore, we limit our analysis to users for whom we could obtain at least 180 days of activity since their very first {\mfp} tweets. Finally, we constrain the analysis to those users who have tweeted at least a month prior to their first {\mfp} tweet (to ensure what we capture is the very first {\mfp} tweet). This step is necessary because Twitter only allows access to the most recent 3,200 tweets, which can cover a variable time span. These steps leave us with 3,169 users, and their 4,794,071 tweets. For a tweet to be considered an {\mfp} tweet in this paper, we also required the ``source'' metadata to indicate that it was generated by {\mfp}, in addition to the presence of the {\hashmfp} hashtag. 

The {\mfp} app generates several types of structured messages in tweets (Table~\ref{table:mfp_auto}). Messages are forwarded to Twitter immediately after an {\mfp} message is generated for every social opt-in user. More specifically, health status updates to SAS include exercise records, weight loss, entries to a fitness diary, addition to on-site community blog, and continuous usage for several consecutive days on {\mfp}. For instance, a tweet example in Table~\ref{table:mfp_auto} reveals the duration and calories burned for a given workout session, and another example shows a user accessed {\mfp} for certain consecutive days. Regular expressions are utilized to extract useful information from this data, including the {\mfp} user names and the amount of calories burned.\footnote{For example, to identify workout tweets containing the amount burned and the workout length, the following expression is utilized:  (({\textbackslash}d*,?{\textbackslash}d+({\textbackslash}.{\textbackslash}+)?){\textbackslash}s(calorie|kilojoule)s?.*({\textbackslash}d+){\textbackslash}smin(ute)?s.*)|
(({\textbackslash}d+){\textbackslash}smin(ute)?s?.*{\textbackslash}s({\textbackslash}*,?{\textbackslash}d+({\textbackslash}.{\textbackslash}d+)?){\textbackslash}s(calorie|kilojoule)s?.*)}

\section{Characteristics of Social Opt-in Users}

This section presents characteristics of social opt-in users, which give insight into understanding the long-term engagement of users.

\subsection{Definition on Social Opt-in User}

\begin{figure}[h]
{
	\centering
	\hspace{-3mm}
	\includegraphics[width=1.02\linewidth]{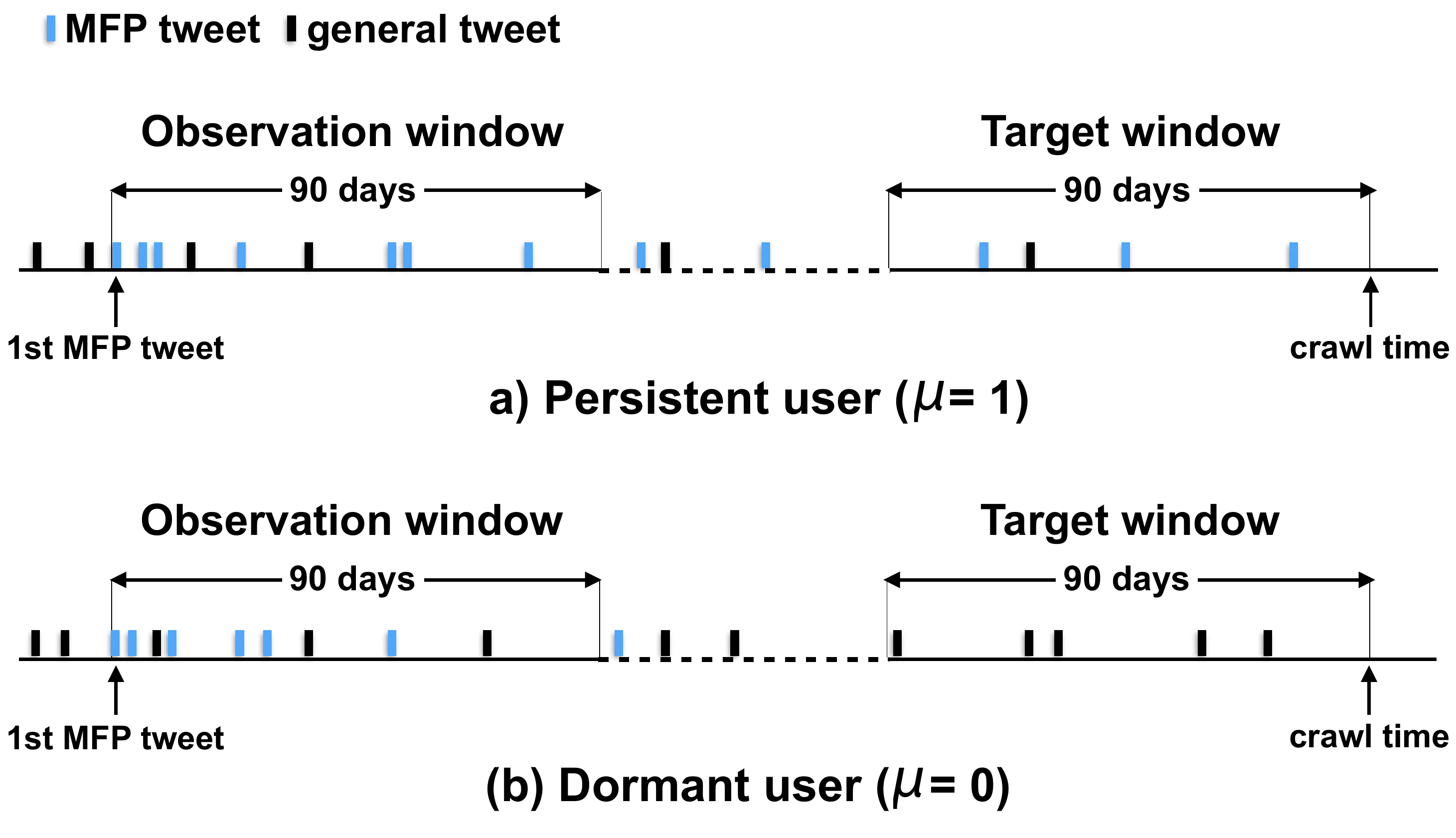}
	\caption{Pictorial description of user persistence ($\mu$)} 
	\label{fig:prediction_definition}
}
\end{figure}

In order to find indicative features of long-term engagement in social sharing, we define a binary variable $\mu$ to represent the persistence of social opt-in. We set the first 90 days since the very first {\mfp} tweet as an ``observation window'' and predict which users continue to share health status on SAS up to the last 90 days of data, which we call a ``target window''. Figure~\ref{fig:prediction_definition} depicts the prediction task, where we set $\mu$=1, if a user posts at least one {\mfp} tweet during the target window and refer such user \textit{persistent}, otherwise we refer a user \textit{dormant} as these users no longer continued to post {\mfp} messages on Twitter and set $\mu$=0. From this definition, we retrieved 2,589 persistent users ($\mu$=1) and 581 dormant users ($\mu$=0), respectively. 
Note that we used a 3 month period in 2014 to identify {\mfp} users on Twitter. Any user who posted a {\mfp} tweet during the crawl period was labeled as social opt-in users. However, the very first {\mfp} tweet of these users could date far back in time, leading to different starting times.

A user can be labeled ``dormant'' for the following reasons. Firstly, a user may no longer use {\mfp}, which is a common case of customer churn. Secondly, a user may have been inactive temporarily, although she may return to the app in a near future. Thirdly, a user may have disabled the auto sharing option while still engaging in the app. All of these cases will result in `dormant' according to Figure~\ref{fig:prediction_definition}. Since we cannot easily distinguish one case from another only from observing Twitter logs, the prediction task focuses on user engagement in the sharing activity but not on retention of the fitness app itself. A measure of persistent social sharing is important nonetheless, because persistent sharing means one is successfully retained to the app.

\subsection{Characteristics of Social Opt-in Users}\label{sec:char_fitness}

\begin{figure*}[t!]
{
	\centering
	\hspace*{-6mm}
	\includegraphics[width=0.5\linewidth]
{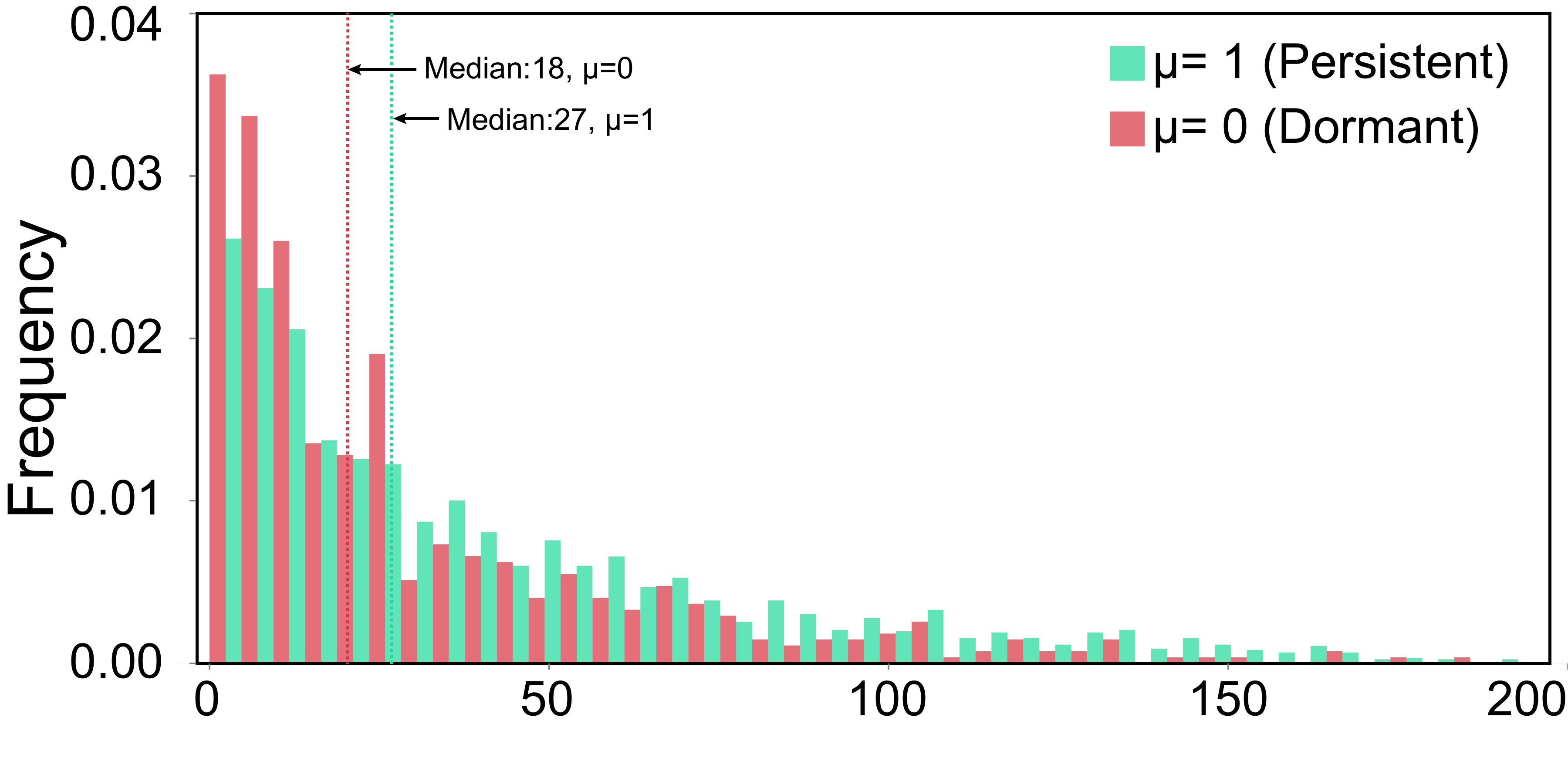}
	\includegraphics[width=0.5\linewidth]{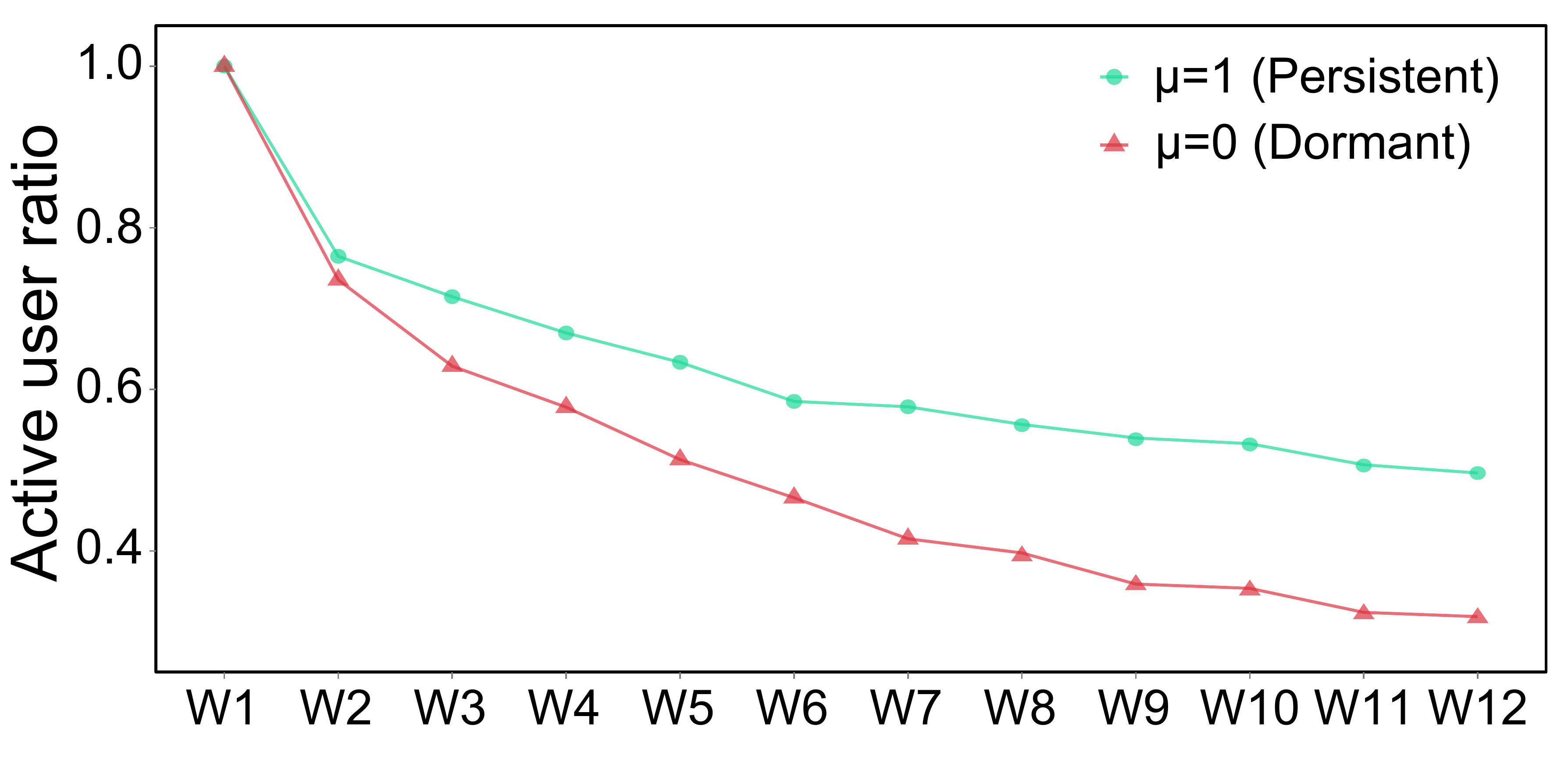}
	{\centering
	(a) Histogram of {\mfp} tweets~~~~~~~~~~~~~~~~~~~~~~~~~~~~~~~~~~~~~~~~~~~~~~~~~~~~~~ (b) Number of active users per week\\}
	\vspace*{1mm}
	\caption{Different amount of engagements of social opt-in users. (a) present histogram of {\mfp} tweets on first 90 days, and (b) displays active user ratio over first 12 weeks.} 
	\label{fig:trends}
}
\end{figure*}

Whether persistent or dormant, social opt-in users had 334 friends on average (i.e., those who follows a user) and 208 followers, with medians of 175 friends and 84 followers, respectively. Unlike follower count that is similar to the global average, friends count is prominently higher than the reported average of 102.\footnote{\url{http://www.beevolve.com/twitter-statistics}} This could be because the Decahose access reaches a sample of tweets, but not users, and hence active users can be overrepresented. Alternatively it could simply be due to changes over time since the reference statistics was compiled.

Users showed a wide range of engagement levels with {\mfp}. Figure~\ref{fig:trends}(a) displays the histogram of {\mfp} tweets in the observation window (i.e., the first 90 days of {\mfp} sharing). The most prolific user posted 194 {\mfp} tweets (i.e., 2.2 MFP tweets per day), whereas 27\% of users posted no more than 10 {\mfp} tweets over the same duration. Figure~\ref{fig:trends}(b) shows the fraction of active users as a function of time, where active user base decreases week by week. Note that those who later become dormant show a rapid drop rate as one might expect. Interestingly up to a quarter of dormant users still remain active after week 12. This means retention rate of social sharing users is much higher than for a typical mobile application. Such tendency of slow churn may be a key characteristic of social opt-in users, as they might feel social accountability and responsibility for their previously publicized commitment, and try to share health update several more times before they quit.

\subsection{Twitter Usage of Social Opt-in Users}

Subsequent to the general characteristics of social opt-in users, we also investigate the set of topics these users engage in on Twitter. Because Twitter is an interest-based social network (i.e., people explicitly follow streams of other people that interest them), one's interest on certain topics will influence how one builds their social network. Below we describe two approaches to analyze the topical interests of users.

We firstly analyze hashtag usage, which can represent topical interest of users. In order to capture how diverse one's hashtags are, we compare both the Shannon's entropy and the number of hashtags between persistent user and dormant user. To test for statistical significance, we utilize the Mann-Whitney U test which does not require normality assumption. Not only dormant users tend to employ a larger number of hashtags than persistent users (34.54 and 36.95, p$<$0.05), but also their entropy in topic diversity is higher (1.88 and 2.06, p$<$0.05). This finding suggests that persistent users are more focused and narrow in their topical interests of Twitter. 

Next we checked how persistent and dormant users post health-related tweets except for {\mfp} tweets. To investigate it,we relied on crowdsourcing. We created a job on CrowdFlower\footnote{\url{http://www.crowdflower.com/}} and provided crowd workers with the screen name, account name, self-description, and five randomly chosen tweets for each user to be labeled. To ensure the test is non-trivial, we only used non-{\mfp} tweets for tagging. Workers were asked the following question: {Is this Twitter user interested in health?} Each question was answered by at least three human coders, aggregated via majority voting. The tagging task yielded moderate agreement rates of 0.577 based on the Fleiss's Kappa value. In case the hired coders did not reach an agreement, the authors manually inspected the user to be labeled and chose the correct answer. We find that 1,368 users out of 3,170 users to have noticeable interest in health. We test whether persistent users have a larger number of users who are interested in health by conducting proportion test with continuity correction. As a result, we find persistent users are more likely to express interest in health on Twitter than dormant users (0.441 and 0.337, p$<$0.05). 

The above findings collectively suggest that having a focus on health topics on Twitter could be an indicator of persistent engagement in fitness applications through social sharing.

\section{Prediction of Persistent Engagement}

We aim to distinguish persistent and dormant users by employing various features from Twitter and {\mfp}. The prediction task is to classify users and determine which set of users, based on the first 90-day logs of {\mfp} (i.e., observation window), will continue to share health status based on the most recent 90-day logs (i.e., target window).

\begin{table*}[t]
{
\centering
\begin{tabular}{llrlrrrll}  
\multirow{2}{*}{Type} & \multirow{2}{*}{Predictor} & \multirow{2}{*}{Coef} & \multirow{2}{*}{SE} & \multirow{2}{*}{p-value} & \multicolumn{2}{c}{Mean value} & \multirow{2}{*}{Description}\\
& & & & & $\mu$=0 & $\mu$=1 & \\\hline 
 & (intercept) &-0.099 & 0.074 & & \\\hline
Twitter & \textit{tweet\_count} & -0.048 & 0.030 & & 180.2 & 163.2 & Number of general tweets\\
Profile & \textit{following} & -0.072 & 0.047 & & 337.5 & 332.6 & Number of friends on Twitter \\
 & \textit{followers} & 0.107 & 0.051 & & 208.2 & 208.1 & Number of followers on Twitter \\
 & \textit{favorite} & -1.982 & 0.488 & *** & 0.052 & 0.031 & Fraction of general tweets favorited \\
 & \textit{retweet} & -0.919 & 0.852 & & 0.022 & 0.016 & Fraction of general tweets retweeted\\\hline
Fitness & \textit{exercise} & 0.276 & 0.037 & *** & 12.51 & 18.72 & Number of workout tweets \\
Activity & \textit{blog} & -0.061 & 0.031 & & 0.234 & 0.193 & Number of blog update tweets \\
 & \textit{usage} & 0.083 & 0.043 & & 0.055 & 0.163 & Number of ``consecutive usage'' tweets\\
 & \textit{weightloss} & 0.007 & 0.033 & & 4.701 & 5.336 & Number of weightloss tweets\\
 & \textit{diary} & 0.071 & 0.036 & & 12.43 & 16.89 & Number of diary update tweets\\
 & \textit{calorie\_mean} & -0.036 & 0.026 & & 1458.6 & 649.0 & Mean of the burned calories\\
 & \textit{extime\_mean} & 0.055 & 0.033 & & 1.751 & 2.040 & Mean of the workout length\\
 & \textit{lost\_weight} & 0.008 & 0.032 & & 16.57 & 18.16 & Total amount of lost weight\\
 & \textit{autoscaled} & -0.044 & 0.029 & & 0.286 & 0.260 & Fraction of auto-scaled over weightloss tweets\\
 & \textit{mfp\_adopt\_date} & 0.008 & 0.003 & ** & 21.31 & 23.88 & Months from first {\mfp} tweet to last tweet\\\hline
Fitness & \textit{mfp\_favorite} & -0.419 & 0.313 & & 0.046 & 0.028 & Fraction of {\mfp} tweets favorited\\
Network & \textit{mfp\_retweet} & 2.106 & 0.955 & * & 0.004 & 0.004 & Fraction of {\mfp} tweets retweeted\\
 & \textit{mfp\_reciprocal} & 0.066 & 0.027 & * &  0.343 & 0.440 & Number of reciprocal friends using {\mfp}\\
 & \multirow{2}{*}{\textit{fitness\_reciprocal}} & \multirow{2}{*}{1.056} & \multirow{2}{*}{0.415} & \multirow{2}{*}{*} & \multirow{2}{*}{0.034} & \multirow{2}{*}{0.048} & Fraction of reciprocal friends who have \\
 & & & & & & & fitness terms on their Twitter bio\\\hline

Model $\chi^2$ & & 265.37 & & *** \\\hline
\multicolumn{8}{r}{*:p$<$0.05, **:p$<$0.01, ***:p$<$0.001} \\
\end{tabular}
\vspace{-2mm}
\caption{Fitted values of standardized logistic regression model for predicting persistent social opt-in user. We observed three months from the initial {\mfp} usage to get features. }
\label{table:result_twitter_based}
}
\end{table*}

\subsection{Basic Setting}

The data contained 2,589 persistent users ($\mu$=1) and 581 dormant users ($\mu$=0). In order to construct a balanced set for the prediction task, we up-sampled from the dormant users and constructed a 5,178 user set allowing repetition. This step enforces a 50-50 prior probability for both $\mu$=1 and $\mu$=0, which makes the classification model meaningful. Without a balanced set, a model would overly be in favor of persistent users. While we do not claim that the real prior probabilities are 50-50, the re-weighing step controls the sample size and helps us identify important factors of persistent engagement.

We now describe the prediction features. Twitter data contain a wide range of features that represent user traits and fitness activities. The first set of features, which are called \textit{Twitter Profile} category in Table~\ref{table:result_twitter_based}, describe the general characteristics of users that are independent of {\mfp}. These features include the number of total days one was active on Twitter as well as the number of followers and friends one has. In counting \textit{tweet\_count}, \textit{favorite}, and \textit{retweet}, we only consider non-{\mfp} tweets (i.e., excluding {\mfp}'s auto-generated tweets). 

The second set of features, grouped as the \textit{Fitness Activity} category, represents the user's activities of {\mfp} portrayed on Twitter. On top of the date when one first shared an {\mfp} tweet (\textit{mfp\_adopt\_date}), they refer to tweets generated by the {\mfp} application of the types shown in Table~\ref{table:mfp_auto}. From tweets of the appropriate type we also extracted the mean amount of calories burned by that user and the amount of exercise time, total lost weight, as well as the fraction of weight loss tweets logged via smart scales, if any exists.\footnote{{\mfp} provides functions to automatically post the weights of users through various commercial WiFi enabled body scales such as the Withings Smart Body Analyzer.}

The last set of features, grouped in the \textit{Fitness Network} category, represents one's social network related to fitness and social support activities a user received from that network on Twitter. The features in this category include the fraction of {\mfp} tweets that are retweeted or favorited. We also examine the effects of {\mfp} network by measuring the number of reciprocal\footnote{The ``friendship'' links on Twitter are, by default, uni-directional and may not be reciprocated.} friends who also use {\mfp}. 
In order to identify peers who are interested in fitness in general (without necessarily sharing fitness-related tweets), we consider reciprocal friends who explicitly mention `health' or `fitness' in their Twitter bio and call this feature \textit{fitness\_reciprocal}. This simple approach can capture whether one is interested in health with high precision. We ensure that these friends are not media sources by excluding those who have more than 10,000 followers by applying the same standard to decide target users.

\subsection{Indicator of Persistent Social Sharing}

In order to determine which features are indicative of persistent social sharing, we utilized standardized logistic regression algorithm and infer the relationship between variables. Table~\ref{table:result_twitter_based} displays the results across three feature categories (i.e., Twitter Profile, Fitness Activity, and Fitness Network). The table presents the coefficient, standard error, statistical significance of each predictor in the logistic regression model. The table also shows the mean value of each variable for persistent and dormant users, as well as the chi-square value of the model that represents the discriminative power of prediction. We make four key observations. 

First, several features in the Fitness Activities category are positively associated with persistent users. Exercise was the only statistically significant feature, while others (including even weight loss) did not show any meaningful relationship. This observation implies that regularly tracking and sharing exercise (rather than posting other forms of updates like diaries) is important for persistent engagement.

Second, we investigate whether any characteristic of social media usage (i.e., Twitter Profile category) is related to persistent users. Users who have a larger fraction of favorited tweets tend to be dormant, which may be that frequent favorites on general tweets make certain users more self-conscious and aware of their audience on Twitter. This finding is related to a previous work that noted perceived audience can affect sharing of personal health behavior~\cite{newman2011s,teodoro2013fitter}.

Third, the fraction of reciprocal friends on fitness and {\mfp}~(i.e., \textit{fitness\_reciprocal}, \textit{mfp\_reciprocal}) have positive effects on persistent user. In contrast, receiving favorites on {\mfp} tweets (i.e., \textit{mfp\_favorite}) did not have any impact on persistent usage. Having an appropriate audience (i.e., having reciprocal friends on fitness and {\mfp}) could be more important than getting explicit support to maintain sharing fitness activities. \textit{mfp\_retweet} is also found to be significant, but it displays the opposite effect with the influence of individual variable. We observed the sign of the effect \emph{reverses} after including the fraction of retweeted tweets (\textit{retweet}) as a feature. Hence, the directionality of the effect of \textit{mfp\_retweet} is potentially due to linearity effects and should be interpreted with caution. 

Fourth, we observe people who start to share {\mfp} tweets earlier than others (i.e., early adopters) are more likely to continue sharing as seen from the \textit{mfp\_adopt\_date} feature. This indicates that early adopters of {\mfp} may have distinct characteristics compared to general users.

Subsequent to findings on features indicative of persistent social sharing, we investigated the importance of individual variables in the prediction task. The logistic regression model in Table~\ref{table:result_twitter_based} can explain which predictor is associated with the target variable after controlling any effects of the remaining variables. However, statistical significance of a predictor does not guarantee the predictive power for a given target variable. Therefore, we additionally utilized the Random Forest model, an ensemble method constructing a number of decision trees and measured variable importance for predicting the persistence in social sharing. In order to quantify the predictive power of individual feature, we calculated the mean decrease in Shannon's entropy of each variable. The entropy measures node impurity, thus a larger decrease of the value indicates that the variable plays an important role in the partitioning data.

\begin{figure}[t]
{
	\centering
	\includegraphics[width=1.05\linewidth]{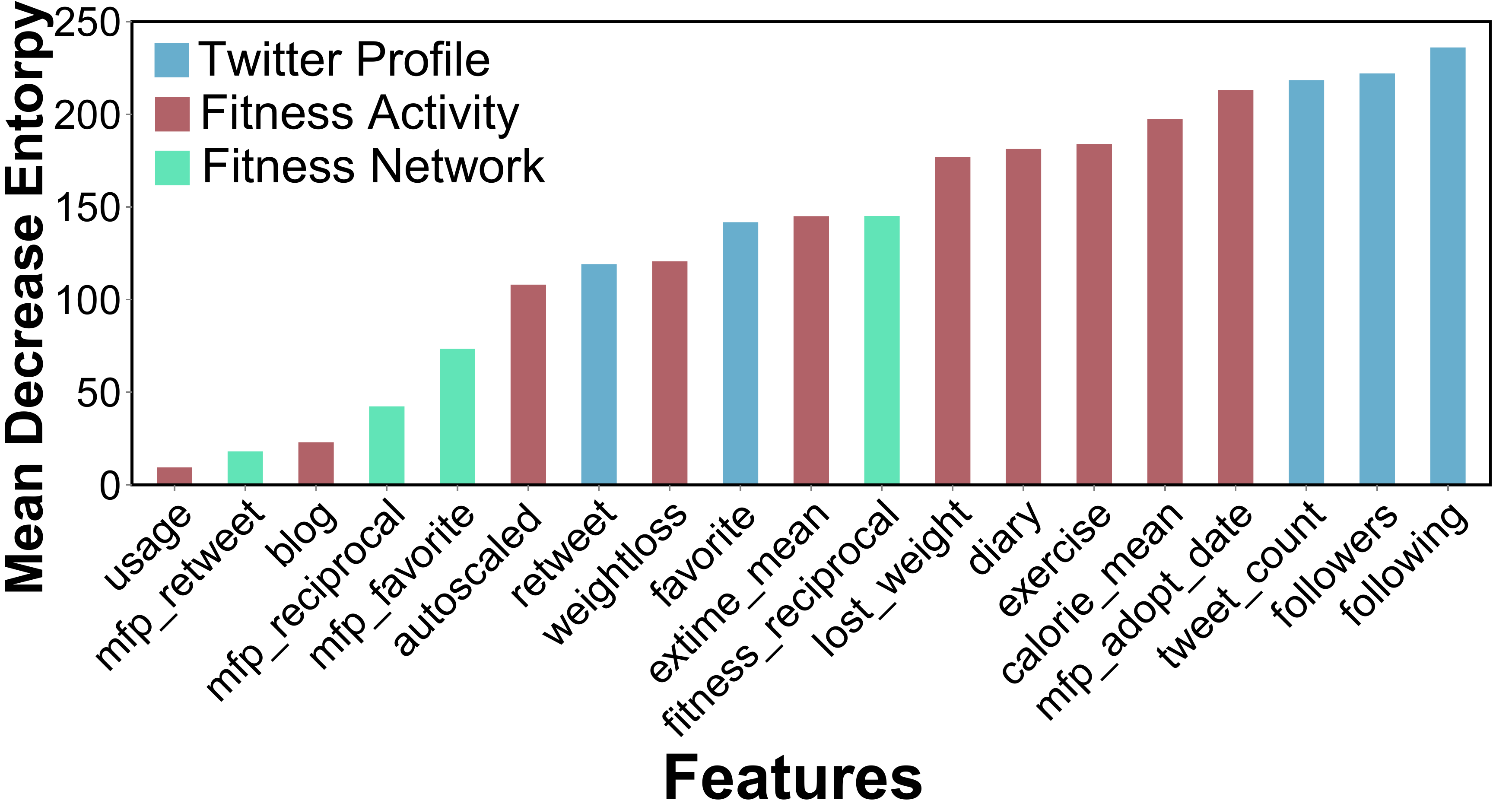}
	\vspace{-4mm}
	\caption{Variable importance measured by the Shannon's entropy for predicting persistent social sharing.}
	\label{fig:variable_importance}
}
\end{figure}

Figure~\ref{fig:variable_importance} displays the variable importance across all 19 features. The figure shows that several features in the Twitter Profile category (\textit{followers}, \textit{following}, \textit{tweet\_count}) have the highest discriminative power, against those features in the Fitness Activity and Fitness Network categories. Among them, interestingly, features based on the general Twitter usage (\textit{followers}, \textit{following}, \textit{tweet\_count}) yield high predictive power for persistent social sharing. This implies that persistent users may have characteristic traits in the general usage of Twitter itself that could distinguish themselves from dormant users. In fact, Table~\ref{table:result_twitter_based} shows that persistent users tend to post fewer tweets (163.2 vs. 180.2) and follow fewer accounts (332.6 vs. 337.5). However, highly ranked features in the Twitter Profile category are not statistically significant in Table~\ref{table:result_twitter_based}, which implies that the differences are basically due to the effects of significant variables in Fitness Activity and Fitness Network. Another finding we make is that several features in the Fitness Activity category are discriminative for persistent social sharing, which reinforces our previous finding that sharing fitness activities frequently on social media is a strong indicator of persistent engagement.

\subsection{Evidence of Seeking Information}

The result of Table~\ref{table:result_twitter_based} highlights the importance of a fitness-related network in persistent sharing of health status. In order to investigate the underlying relationship of Twitter network on persistent social sharing, we now focus on other features that are descriptive in nature such as one's followings list (also called friend list). The followings list determines the tweets from whom a user would like to receive feeds from and is used for various recommendations within Twitter. As mentioned in~\cite{epstein2015nobody}, people can share their personal health status on SAS to get informational support from existing network. We believe seeking information can be represented by the list of following users, so we would like to test whether such following relationship on Twitter have any predictive power on long-term social sharing of fitness applications. 

Does following a particular account and receiving that account's feeds increase (or decrease) the probability of long-term engagement? Table~\ref{table:top.10.discriminative.friends} shows the list of top-10 discriminative accounts sorted by the chi-squared value for persistent and dormant users, respectively. The \textit{Prob} column shows the conditional probability of persistent social sharing when a user follows the given account on Twitter, for a 50-50 balanced sample. The table also shows the number of users who follow such accounts out of 3,170 {\mfp} users in the column denoted \textit{Freq}.

Persistent users ($\mu$=1) were far more likely to follow accounts related to fitness and diet such as AboutCalorie, FitnessNHappier, and AHealthyBody. These accounts appeared frequently among the persistent users' lists. On the other hand, dormant users ($\mu$=0) exhibited a diverse set of interests, where many of them were related to entertainment---such as television shows, music, and movies. It is interesting to observe that certain celebrities such as IAMJHUD (Jeniffer Huddson) and TheRealMikeEpps (Mike Epps) were favored by persistent users, although in general television shows were favored by dormant users.

\begin{table}[t]
{
\hspace*{-3mm}
\centering \frenchspacing
\footnotesize
\begin{tabular}
{llc||llc}
\multicolumn{3}{l||}{\textsf{\textbf{Increase prob. for $\mu$=1}}} & \multicolumn{3}{l}{\textsf{\textbf{Increase prob. for $\mu$=0}}} \\
\multicolumn{3}{l||}{\textsf{Account~~~~~~~~~~~~~~~~Prob~~~Freq}} & \multicolumn{3}{l}{\textsf{Account~~~~~~~~~~~Prob~~~Freq}}\\ \hline
AboutCalorie & .81 & 206 & SimonCowell & .36 & 183\\
DaveRamsey & .77 & 142 & Ginofantastico & .30 & 56\\
FitterNHappier & .79 & 125 & JKCorden & .37 & 129\\
HowManyCaloriee & .83 & 89 & HulkHogan & .33 & 57\\
AHealtyBody & .73 & 142 & BetteMidler &.35 & 72\\
IAMJHUD & .72 & 147 & Celebjuice & .34 & 60 \\
JoyceMeyer & .66 & 213 & Vangsness & .34 & 52 \\
TheRealMikeEpps & .79 & 87 & WorkoutHumor & .34 & 52 \\
evernote & .83 & 68 & michkeegan & .34 & 52 \\
JoeManganiello & .83 & 66 & IamStevenT & .35 & 65 \\\hline
\end{tabular}
\vspace{0mm}
\caption{Screen names of Twitter accounts that are discriminative of persistent and dormant user, and their probability for $\mu$=1 and the frequency of incoming links per account.} 
\label{table:top.10.discriminative.friends}
}
\end{table} 

\newpage
\subsection{Is Fitness Activity More Important than Fitness Network?}

Features related to both Fitness Activity and Fitness Network were important for persistent sharing, yet those in the Fitness Activity category had higher predictive power. Is Fitness Activity then more important that Fitness Network for persistence? If so, how important is Fitness Activity for persistent users compared to Fitness Network? In order to answer these questions, we examine the relationship between the Fitness Activity and Fitness Network.

Table~\ref{table:user_type} displays the proportion of persistent users across two of the significant axes: exercise-related {\mfp} tweets and presence of a fitness support network. We consider a user belongs to a `high' exercise group if one posted more exercise {\mfp} tweets than the median value across all users. Otherwise we consider a user to be in a `low' exercise group. In addition, the presence of Fitness Network is considered to be `high' if a user falls into the upper median of either \textit{mfp\_reciprocal} or \textit{fitness\_reciprocal}, which are two predictive features of persistent sharing. Otherwise, we consider a user has `low' presence of Fitness Network. 

\begin{table}[ht]
{
\tabulinesep=1.2mm
\centering \frenchspacing
\begin{tabu}{c|c|cc|c|}
\multicolumn{2}{c}{} & \multicolumn{3}{c}{Exercise-related {\mfp} tweet} \\\hhline{~~|--|-}
\multicolumn{2}{c|}{}  & Low  & High & All\\\hhline{~-|--|-|}
\multirow{3}{1.3cm}{ Fitness Network}& Low & 499 (38.2\%) & 621 (56.4\%) & 1120\\\hhline{~~|~~|~|}
& High & 624 (46.8\%) & 845 (58.7\%) & 1469\\\hhline{~-|--|-|}
& All & 1123 & 1466 & 2589\\\hhline{~-|--|-|}
\end{tabu}
\vspace{0mm}
\caption{Number of persistent users across exercise on {\mfp} and presence of Fitness Network. Percentages within parenthesis indicate fraction of persistent user over total social opt-in user per each group.} 
\label{table:user_type}
}
\end{table} 

Table~\ref{table:user_type} shows the results where each quadrant shows the raw and the proportion of persistent users who belong to that area. Between the amount of exercise {\mfp} tweets and the presence of fitness network, posting a higher number of exercise tweets is more powerful to discriminate persistent users as we found in Figure~\ref{fig:variable_importance}. The presence of a fitness network generally increases the probability of persistent usage. However, fitness users tend to be dormant if they had posted fewer number of exercise tweet even when they possess a sizable fitness network. In a similar vein, social opt-in users tend to be persistent with absence of fitness network but with a higher number of exercises. In the group of 621 users in Table~\ref{table:user_type} who posted many exercise tweets and owned a small fitness network, we observe that 22 persistent users had zero-sized fitness networks, and 15 of them further do not post any general tweets. They may not need support of fitness network on Twitter. We hypothesize that some {\mfp} users utilize social sharing for archiving purpose, which requires further investigations in future work.

\section{Discussion and Conclusion}

It is well known that many users of fitness apps discontinue with their usage after a few weeks of trial, which is commonly known as a \textit{user retention} problem. This relatively low retention rate of health and fitness app users has drawn the attention of research community~\cite{leslie2005engagement}. Promoting social sharing of users has been considered to be one of many important strategies to boost their long-term engagement level. Several researchers have studied different sharing behavior of health and fitness-related activity on social media outlets. Some studies analyzed the type of content and motivations for
sharing health-related activity on social media outlets  ~\cite{newman2011s,teodoro2013fitter} while others conducted a qualitative analysis of the ways that people think
about with whom and how to share different types of information as they pursue social goals related to their
personal health, including emotional support, motivation, accountability, and advice ~\cite{munson2010happier,newman2011s}. One important issue that has been neglected 
within the research community, though, is the question of \textbf{\textit{what are characteristics of users who share health and fitness-related activity on social media outlets over a long period of time compared to those short-lived users?}} We believe that it is important to properly address this question especially when it is unknown how the overall effectiveness of social sharing of health-related activity is related to the future engagement level of fitness apps. In this paper, through a quantitative analysis of 3,169 Twitter users, we have found some partial answers to this question. Below we summarize our findings on persistent social sharing of health and fitness-related activity and their implications.

\noindent \textbf{1. Persistent users are more likely to use Twitter in a health-oriented way:} We observe that dormant users tend to be interested in a more diverse set of topics while 
persistent users tend to be interested in a more narrow set of topics that are usually related to fitness or health. Thus, we speculate whether this variety of interest topics that is normally shown by dormant users could inhibit these from gaining appropriate audiences interested in health and fitness status updates. 
We believe that this is particularly true with Twitter as it is an information-based social network and the topical diversity of user's Twitter feed can have a certain impact on the 
richness of her network in Twitter. Possessing a fitness-related social network has been shown to be effective for getting positive effects~\cite{munson2010happier,newman2011s}, hence a topical diversity of their audience might prohibit fitness app users to share their health status, leading them to be dormant. Our work, though, cannot provide a definitive conclusion on such causal relationships since we did not test the direct effect of having diverse interests on the formation of a fitness related network and impact on user's emotions. Future work can further delve into a better understanding of how expressing diverse interests on Twitter would affect persistent sharing of health and fitness status.

\noindent \textbf{2. Sharing exercise related tweets is positively correlated with the long-term engagement:} We observe that persistent and dormant users differ in the type of content that they share on social media. More specifically, content related to different physical exercises is found to be prevalent in tweets shared by persistent users. Several studies have found that physical exercise is important to maintain one's healthiness~\cite{Elfhag:2005wz}, and the most recent one found that exercising can even affect one's mental health (e.g., anxiety, depression)~\cite{reisculotta15aaai}. Our finding corroborates the importance of exercising regularly and those who do tend to show a long-term engagement with fitness apps.

\noindent \textbf{3. Having a mutual friendship with those users who share health and fitness related content on social media or heavily using {\mfp} is positively correlated with persistent social sharing:} The importance of a ``fitness network'' for persistent social sharing is a key takeaway of this paper. We observe the importance of one's social network to persist with sharing of her health and fitness status updates. While the explicit support of favorites one receives on tweets related to health and fitness activity was not necessarily linked to persistent social sharing, the fraction of {\mfp} reciprocal friends (\textit{mfp\_reciprocal}) and fitness friends on Twitter (\textit{fitness\_reciprocal}) were found to be indicative of persistent social sharing. These findings imply that \textit{sustaining a support fitness network of reasonable size} becomes important for long-term engagement, rather than just receiving explicit responses from peers in their social network. Having a network of health interests may help users enjoy social support, which was found to be a key driver for a long-term engagement of users to fitness activities~\cite{consolvo2006design} and in health forums~\cite{vydiswaran2014user}.

\noindent \textbf{4. Frequently receiving favorites on tweets is negatively correlated with long-term social sharing:} One interesting finding is that users who have a larger fraction of favorited tweets related to health and fitness tend to be dormant. Since dormant users can result from several reasons, there can be many explanations of this particular finding. One possible explanation is that considering the delicacy of sharing health and fitness status via social-media outlets, receiving favorites too frequently will make her conscious of her audience on Twitter. Future work could potentially analyze how those popular social media users feel when they receive favorites on their tweets, and how this would affect their long-term engagement.

\subsection{Towards a Better Design of Fitness Apps}

Our findings lead to a couple of guidelines that could be useful during the design of fitness apps with the goal of motivating fitness app users better. 

\subsubsection{1) Introduction of a rewarding program}
Our finding on the positive correlation between physical exercise and social sharing persistence highlights the potential improvement of user engagement for fitness apps by introducing an incentive-reward program. In other words, fitness apps could devise a reward or badge-based gamification features that can motivate their users better, where some highly correlative user activities shown in Table~\ref{table:result_twitter_based} might get extra incentives by offering special badges to those users who continuously engage with these activities~\cite{fritz2014persuasive}. Alternatively, those behaviors that bring benefit to the whole fitness eco-system may be rewarded with strong incentives~\cite{robertson2009rethinking}. For instance, posting a specific number of physical exercise related tweets, which is shown to have the highest predictive power in Table~\ref{table:result_twitter_based}, should be incentivized. By encouraging their users to engage more with the most predictive features, fitness apps may potentially enjoy a higher user retention rate while empowering their users to be more effective with health and fitness goals. The true effectiveness of using badges as a incentive-rewarding program albeit needs to be tested in practice especially when it is reported that offering badges might lead to some negative effects~\cite{munson2012exploring}.

\subsubsection{2) Connecting peers over fitness application}
Our study shows that persistent users tend to own a strong health and fitness-oriented Twitter network. Persistent users not only have reciprocal friends who are active in fitness and {\mfp} users, but also tend to follow fitness-related popular accounts and seek information. Grounded by the findings, fitness apps could potentially adopt certain features that would help users connect better with fitness-related Twitter accounts. Since ``Informational Support'' is a primary goal of peer-support network as shown in ~\cite{epstein2015nobody}, the nature of Twitter could easily meet this goal by recommending Twitter accounts that can act as useful sources of health and fitness related information. We do not know yet the temporal order of long-term engagement and fitness-related network construction. Persistent users are more likely to have strong health and fitness network, but it is also possible that these users built up the supportive group by actively engaging with {\mfp}. To overcome this limitation, future work could investigate this trend by analyzing overall dynamics of following links on Twitter or by conducting in-depth interviews of persistent users.

\subsection{Limitations}

Our work has certain limitations. First there is potential sampling bias. Since the Decahose and is a tweet-based sample, our data are biased toward selecting users who tweet frequently. While this is a common limitation for studies employing data APIs, we also mention that it is non-trivial to define what is a ``good'' uniform user sample on Twitter and the industry standard of ``monthly active users'', as defined through log in events, cannot even be observed by outsiders.

Second is the lack of strong temporal data linkage. Several features such as those in the Profile category were independent of the observation window in Figure~\ref{fig:prediction_definition}, since the Twitter profile is captured from the snapshot of crawling time. Therefore, we cannot be certain of the temporal order of effects of the social network (e.g., \textit{mfp\_reciprocal}, \textit{fitness\_reciprocal}). Persistent social opt-in users might have fitness-related audiences because they have engaged in fitness activities for a long time, rather than the audience leading to long-term usage. In future work, we hope to better understand this temporal order through repeated data collection and user interviews.

\newpage
\subsection{Conclusion and Future Work}

This paper presented a quantitative study of a popular fitness application, MyFitnessPal and investigated the patterns of sharing personal health status on Twitter. We examined which features of the Twitter profile, the fitness activities, and the fitness network are related with persistent social sharing and found that a wide set of features from one's fitness activities and fitness network have positive correlations.

We observed an interesting signal related to persistent social sharing, in that the date when a user first shared a {\mfp} tweet on Twitter (i.e.\ \textit{mfp\_adopt\_date}) was a significant indicator of persistent sharing. Early adopters may have distinct personality related to engagement. Alternatively, starting to engage in an application earlier than others may make them loyal to engage in the app for a long time. We believe that this is worth to investigate in the future, which could give insights to application providers.

We also noticed exceptional Twitter usage patterns related to sharing health status. Certain users did not have any general tweets. That is, those users utilized Twitter only for {\mfp}. While these users can be simply less active on Twitter, it is probable that these users created a separate Twitter account exclusively to broadcast their {\mfp} tweets. We found 97 users who have only {\mfp} tweets, even though the fraction of persistent users are not significant (79 out of 97). An exciting future direction would be to investigate why those users have dedicated accounts for a fitness app, and how it affects persistent engagement.

Social opt-in users of the {\mfp} application showed a high retention rate; 2,589 out of 3,169 users remained active persistently, which means that they were successfully retained to the app. We do not claim social sharing necessarily increases the probability of retention, because we did not test any direct effects of social sharing under controlled experiments on randomized samples. Nevertheless, an extremely high retention rate seen from persistent users may be related in part to the positive effects of social sharing. Similar finding has been shown by one research group, where integrating a health application within a social network like Facebook increased the retention rate~\cite{munson2010happier}. It is also possible that only more committed users choose to opt-in to social sharing in the first place. In future works, characteristics of social opt-in users and the effects of social sharing on user retention of the fitness application could be investigated via controlled experiments and observational studies with randomized conditions on fitness application.

Beyond investigation of persistent social sharing on Twitter, a natural extension of this work would be to analyze characteristics of persistent social sharing on other social media. Facebook can be a good candidate for looking into persistent sharing of health status, because it is the leading social service where over billions of people engaged. Although some negative effects are reported when sharing a health status on Facebook~\cite{newman2011s}, it would be interesting to know the characteristics of usage and network patterns of Facebook users who share personal fitness status in the long run. Instagram can be another good place to look into. This service is an image-based social network, so the users may show different usage and network patterns compared to general online social network such as Twitter and Facebook. It would be interesting to know why users share images related to fitness and health statues, and how usage and network patterns are related to persistency in social sharing.

\section{Acknowledgements}

We want to thank the anonymous reviewers for their extremely useful comments and suggestions on this paper. This work was partly supported by the BK21 Plus Postgraduate Organization for Content Science of Korea and Institute for Information \& communications Technology Promotion (IITP) grant funded by the Korea government (MSIP) (R0184-15-1037, Development of Data Mining Core Technologies for Real-time Intelligent Information Recommendation in Smart Spaces).

{
\bibliographystyle{SIGCHI-Reference-Format}
\bibliography{references}
}
\end{document}